\UseRawInputEncoding

\documentclass[%
 reprint,
superscriptaddress,
 amsmath,amssymb,
 aps,
]{revtex4-2}

\usepackage{graphicx}
\usepackage{dcolumn}
\usepackage{bm}
\usepackage{colortbl}
\usepackage[table,xcdraw]{xcolor}
\usepackage{caption}
\usepackage{booktabs}
\DeclareCaptionLabelFormat{natcom}{\textbf{Fig. #2 }}
\DeclareCaptionLabelFormat{nctable}{\textbf{Table 1 }}
\DeclareCaptionLabelSeparator{vert}{$\vert$}
\usepackage[normalem]{ulem}
\usepackage{lineno}

\usepackage{ulem}

\newcommand{\remove}[1]{}

\begin{document}


\title{Enhancing the efficiency of polariton OLEDs in and beyond the single-excitation subspace}
\author{Olli Siltanen}
\email{olmisi@utu.fi}
\affiliation{Department of Mechanical and Materials Engineering, University of Turku, FI-20014 Turun yliopisto, Finland}
\author{Kimmo Luoma}
\affiliation{Department of Physics and Astronomy, University of Turku, FI-20014 Turun yliopisto, Finland}
\author{Andrew J. Musser}
\affiliation{Department of Chemistry and Chemical Biology, Cornell University, Ithaca, NY 14853, USA}
\author{Konstantinos S. Daskalakis}
\email{konstantinos.daskalakis@utu.fi}
\affiliation{Department of Mechanical and Materials Engineering, University of Turku, FI-20014 Turun yliopisto, Finland}

\date{\today}

\begin{abstract}
    Organic light-emitting diodes (OLEDs) have redefined lighting with their environment-friendliness and flexibility. However, only 25 \% of the electronic states of organic molecules can emit light upon electrical excitation, limiting the overall efficiency of OLEDs. Strong light-matter coupling, achieved by confining light within OLEDs using mirrors, creates hybrid light-matter states known as polaritons, which could ``activate" the remaining 75 \% electronic triplet states. Here, we study triplet-to-polariton transition and derive rates for both reverse inter-system crossing and triplet-triplet annihilation. In addition, we explore how the harmful singlet-singlet annihilation could be reduced with strong coupling.
\end{abstract}


\maketitle


Organic light-emitting diodes (OLEDs) offer several advantages over traditional lighting alternatives. One key aspect is their versatility in design and form; OLEDs are incredibly thin, lightweight, and flexible, allowing for innovative lighting solutions and high-definition displays~\cite{Forrest2004}. However, due to spin statistics, electrical injection in molecular materials results in 25~\% of the excitations to populate singlet electronic states, the remaining 75~\% populating triplets. Typically, singlets are favored due to their ability to undergo fluorescence, which is substantially faster compared to phosphorescence, thus reducing the likelihood of losses from exciton-exciton and exciton-polaron collisions \cite{Trindade2023,Tankeleviciute2024}. While optical excitation is experimentally simpler and allows to create more singlets, electrical excitation---despite the problematic triplets---remains practical in real-world applications. The performance of such devices can be enhanced by converting the triplets into singlets or rapidly and radiatively depopulating them~\cite{mischok2023highly,Yoshida2023,Zhao2024}.

Emitters displaying thermally activated delayed fluorescence (TADF) have recently emerged as a class of OLED emitters with high internal quantum efficiency (IQE). In TADF emitters, triplet excitations are efficiently converted into singlets by reverse inter-system crossing (RISC)~\cite{Uoyama2012}. Generally, high RISC rates can be achieved by tuning orbital overlaps or incorporating charge-transfer character into the wave functions of the lowest electronic states~\cite{Diesing2024}. However, such molecular design techniques are very demanding, and they can simultaneously reduce the oscillator strength of the singlet-to-ground state transition, weakening fluorescence. Thus, alternative strategies to achieve high RISC rates without compromising the emitters' ability to emit photons are needed.

Microcavity polaritons have been recently introduced as a system that can resolve the RISC-IQE trade-off. Polaritons, eigenstates emerging from the strong coupling of singlet excitons and cavity photons, can act as artificially ``Stokes-shifted" singlet states~\cite{Sanvitto2016,Feist2018,Hertzog2019}. In the context of OLEDs and beyond, this means that by utilizing straightforward cavity designs~\cite{Martinez-Martinez2019,Bhuyan2023,Mukherjee2023}, the emitters inside a cavity can exhibit high RISC rates \textit{and} high IQE, resulting in optoelectronic devices combining simple architectures and superior performance. While preliminary experimental results have been reported~\cite{Stranius2018,Eizner2019,Berghuis2019,Yu2021,Ye2021,Nanoph2023}, the theoretical models are quite rudimentary, limiting our understanding on how polaritons interact with these molecular processes. This is a critical bottleneck that hinders efficient triplet harvesting in actual OLED devices.

\begin{figure*}[t!]
  \includegraphics[width=\textwidth]{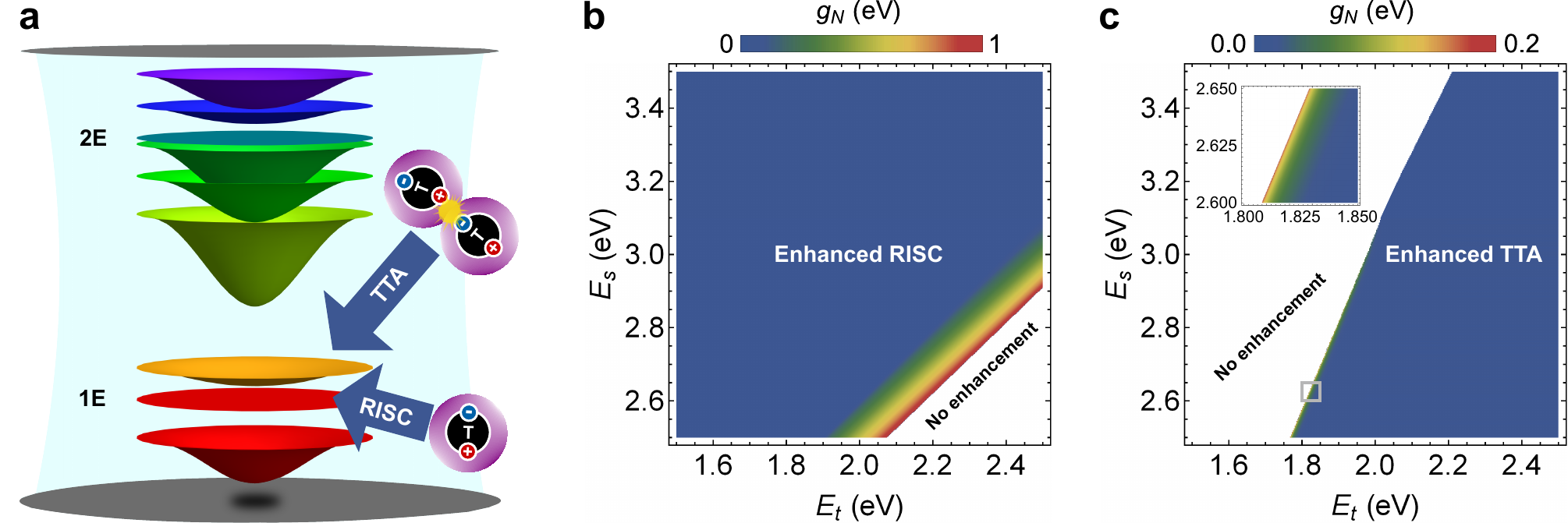}
  \caption{
  \textbf{ Overview of the study and enhancement maps. a,} Schematic picture of polaritons in the 1- and 2-excitation subspaces (1E and 2E), RISC, and TTA. The different surfaces are the new eigenenergies as functions of in-plane momentum, while the spheres represent localized triplet excitons. \textbf{b,} Singlet and triplet energies of enhanced RISC under strong coupling. \textbf{c,} Singlet and triplet energies of enhanced TTA under strong coupling. In \textbf{b} (\textbf{c}), the heat map gives the coupling strength $g_N$ at which the polaritonic RISC (TTA) rate equals the bare-film rate. In the white areas, polaritonic RISC and TTA cannot be enhanced with any $g_N$. The inset in \textbf{c} is a magnification of the highlighted area. The parameters in \textbf{b} and \textbf{c} are $N=10^{10}$, $n_\textit{eff}=2$, $m=1$, $L_c=100$ nm, $k_\Vert=0$, $T=293$ K, $\lambda_s=0.10$ eV, and $\lambda_\pm=0.79$ eV.
  }
  \label{fig:schematic}
\end{figure*}

In this work we introduce a theoretical model for polaritonic OLED processes not restricted to the single-excitation subspace, allowing us, for the first time, to explore RISC together with triplet-triplet and singlet-singlet annihilation (TTA and SSA). In particular, we derive rates for both polaritonic RISC and TTA, an \textit{intermolecular} mechanism of triplet-to-singlet transition, by using the Marcus theory of electron transfer and Fermi's golden rule. The system is illustrated in Fig.~\ref{fig:schematic}a. By scanning the parameter space of our model, we construct ``enhancement maps" for both RISC and TTA; Figs.~\ref{fig:schematic}b and c give the singlet and triplet energies and minimum coupling strengths required to enhance RISC and TTA, respectively. In addition, we analyze the effect of the number of molecules and study how SSA could be reduced with strong coupling. Finally, we apply our model to six molecules previously studied under strong coupling~\cite{Stranius2018,Eizner2019,Berghuis2019,Yu2021,Ye2021,Nanoph2023}.

\newpage

\section*{Results}

\subsection*{The system}

We consider a system of $N$ identical organic molecules carrying 0--2 excitations in total, all coupled to a single cavity mode. The system can be described by the Tavis-Cummings Hamiltonian~\cite{Bhuyan2023}
\begin{equation}
\begin{split}
    &H=\sum_{n=1}^NE_s|S_n\rangle\langle S_n|+\sum_{n=1}^NE_t|T_n\rangle\langle T_n|+E_c\hat{a}^\dagger\hat{a}\\ &\hspace{20pt}+\sum_{n=1}^Ng_1\Big(|S_n\rangle\langle G_n|\hat{a}+|G_n\rangle\langle S_n|\hat{a}^\dagger\Big),
\end{split}
\label{eq:HTC}
\end{equation}
where we have used the rotating-wave approximation, $S_n$, $T_n$, and $G_n$ denote a singlet, triplet, and ground-state exciton at the molecular site~$n$, respectively, $E_{s(t)}$ is the singlet (triplet) energy---for simplicity, we have set the ground-state energy to zero---$\hat{a}^\dagger$ is the creation operator of a photon with the energy $E_c$, $\hat{a}$ is the corresponding annihilation operator, and $g_1=\mu\sqrt{E_c/(2\epsilon_0V)}$ (with $V$ being the mode volume) is the light-matter coupling strength. Finally, the energy of the cavity mode satisfies
\begin{equation}
E_c=\frac{\hbar c}{n_\textit{eff}}\sqrt{\frac{m^2\pi^2}{L_c^2}+k_\Vert^2},
\label{eq:Ec}
\end{equation}
where $n_\textit{eff}$ is the refractive index of the medium, $m\in\mathbb{N}$, $L_c$ is the length of the cavity, and $k_\Vert$ is the in-plane momentum.

Notably, we have assumed the dominance of singlet-cavity mode coupling over all other couplings in Eq.~\eqref{eq:HTC}; The transition dipole moment (TDM) $\mu$ of triplets is typically negligible for non-phosphorescent molecules~\cite{Bhuyan2023}, which allows us to omit strong interactions between the triplet states and cavity mode. Phonon-couplings (and phonons altogether) can be neglected due to polaritons being able to decouple electronic and vibrational degrees of freedom~\cite{Herrera2017}. Finally, the $S$-$T$ couplings $g_{st}$ can be omitted if $|E_s-E_t|\gg g_{st}$, i.e., if the energy required to transition between these states is so high that the perturbation provided by the coupling is insufficient to cause significant mixing. This ensures that the $S$-$T$ coupling terms oscillate rapidly and average out, allowing the singlets and triplets to be treated as effectively decoupled for the purposes of solving the system's eigenstates (polaritons).

\subsection*{Single-excitation subspace}

Diagonalizing the Hamiltonian~\eqref{eq:HTC}, we arrive at the $N$ trivial eigenstates $|T_n\rangle$ in the triplet manifold and the following $N+1$ eigenstates in the singlet-cavity mode manifold with one excitation,
\begin{align}
    |P_+\rangle&=\frac{\alpha^{(1)}}{\sqrt{N}}\sum_{n=1}^N|S_n\rangle\otimes|0\rangle+\beta^{(1)}|\mathcal{G}\rangle\otimes|1\rangle,
\label{eq:UP}\\
    |P_-\rangle&=\frac{\beta^{(1)}}{\sqrt{N}}\sum_{n=1}^N|S_n\rangle\otimes|0\rangle-\alpha^{(1)}|\mathcal{G}\rangle\otimes|1\rangle,
\label{eq:LP}\\
    |D_k\rangle&=\frac{1}{\sqrt{N}}\sum_{n=1}^Ne^{i2\pi nk/N}|S_n\rangle\otimes|0\rangle,k\in[1,N-1].
\label{eq:dark}
\end{align}
$|P_+\rangle$ is the upper polariton (UP) and $|P_-\rangle$ the lower polariton (LP), whereas $\{|D_k\rangle\}$ constitutes the non-emitting exciton reservoir. In the above expressions, the states not explicitly shown are in their electronic ground states and $|\mathcal{G}\rangle$ denotes the global (electronic) ground state. The matching eigenvalues are
\begin{equation}
    E_\pm=\frac{E_s+E_c}{2}\pm\sqrt{g_N^2+\frac{(E_s-E_c)^2}{4}}
\end{equation}
for the polaritons ($+$ for UP and $-$ for LP) and $E_s$ for the dark states. Here, $g_N=\sqrt{N}g_1$. Finally, it is straightforward to show that the parameters $\alpha^{(1)}$ and $\beta^{(1)}$ satisfy
\begin{align}
    |\alpha^{(1)}|^2&=\frac{1}{2}\Bigg(1+\frac{E_s-E_c}{\sqrt{(E_s-E_c)^2+4g_N^2}}\Bigg),\\
    |\beta^{(1)}|^2&=\frac{1}{2}\Bigg(1-\frac{E_s-E_c}{\sqrt{(E_s-E_c)^2+4g_N^2}}\Bigg),
\end{align}
the squares being known as the Hopfield coefficients~\cite{Hopfield1958}.

\subsection*{Two-excitation subspace}

In realistic systems, there can be many excitations present at the same time. Diagonalizing the Hamiltonian~\eqref{eq:HTC} in the presence of \textit{two} excitations, we arrive at the $N(N-1)/2$ trivial eigenstates $|T_mT_n\rangle$ ($m<n$) in the triplet manifold. In the polariton manifold, we use the ansatz
\begin{equation}
\begin{split}
|\psi_l^{(2)}\rangle=&\alpha_l^{(2)}\sqrt{\frac{2}{N(N-1)}}\sum_{m<n}|S_mS_n\rangle\otimes|0\rangle\\
&+\beta_l^{(2)}\frac{1}{\sqrt{N}}\sum_n|S_n\rangle\otimes|1\rangle+\gamma_l^{(2)}|\mathcal{G}\rangle\otimes|2\rangle,
\end{split}
\end{equation}
where $l$ labels the eigenstate ($l=+,0,-$), and the \textit{second-order} Hopfield coefficients satisfy $|\alpha_l^{(2)}|^2+|\beta_l^{(2)}|^2+|\gamma_l^{(2)}|^2=1$. Then, from $H|\psi_l^{(2)}\rangle=\xi_l|\psi_l^{(2)}\rangle$ we get the following system of equations,
\begin{equation}
\begin{cases}
\alpha_l^{(2)}2E_s+\beta_l^{(2)}\sqrt{2}g_1=\alpha_l^{(2)}\xi_l,\\
\beta_l^{(2)}(E_s+E_c)+\big(\alpha_l^{(2)}+\gamma_l^{(2)}\big)\sqrt{2}g_1=\beta_l^{(2)}\xi_l,\\
\gamma_l^{(2)}2E_c+\beta_l^{(2)}\sqrt{2}g_1=\gamma_l^{(2)}\xi_l.
\end{cases}
\end{equation}
Finally, using the normalization condition of the second-order Hopfield coefficients and the approximation $N-1\approx N$, we get
\begin{align}
    \xi_0&=E_s+E_c,\\
    \xi_\pm&=2E_\pm,\\
    |\alpha_l^{(2)}|^2&=\Bigg[1+\Big(\frac{\xi_l-2E_s}{\xi_l-2E_c}\Big)^2+\Big(\frac{\xi_l-2E_s}{\sqrt{2}g}\Big)^2\Bigg]^{-1},\\
    |\beta_l^{(2)}|^2&=\Bigg[1+\Big(\frac{\sqrt{2}g}{\xi_l-2E_c}\Big)^2+\Big(\frac{\sqrt{2}g}{\xi_l-2E_s}\Big)^2\Bigg]^{-1},\\
    |\gamma_l^{(2)}|^2&=\Bigg[1+\Big(\frac{\xi_l-2E_c}{\xi_l-2E_s}\Big)^2+\Big(\frac{\xi_l-2E_c}{\sqrt{2}g}\Big)^2\Bigg]^{-1}.
\end{align}

\begin{figure}[t!]
  \includegraphics[width=\linewidth]{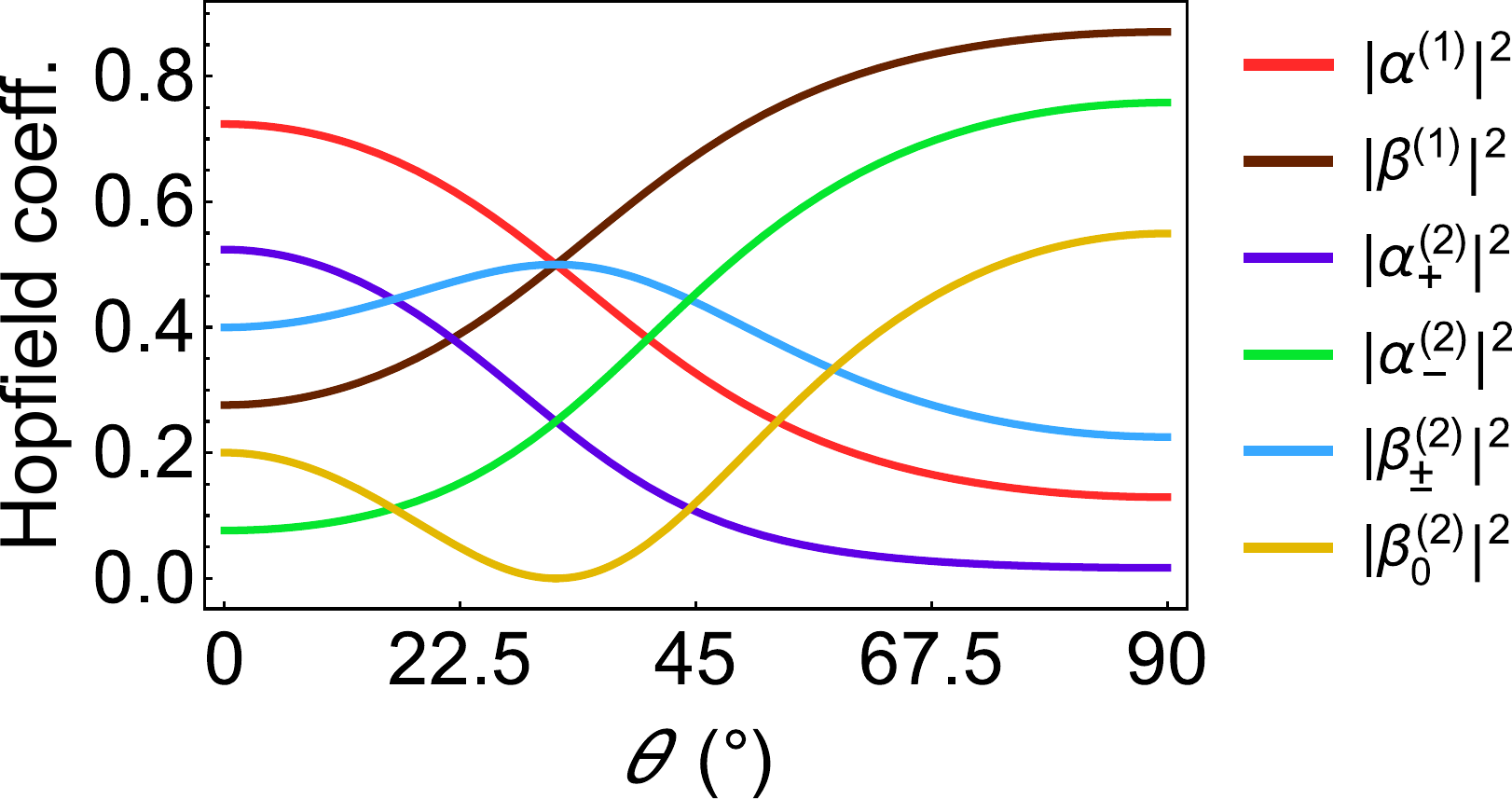}
  \caption{
  \textbf{ The first- and second-order Hopfield coefficients.} The example parameters are $n_\textit{eff}=2$, $m=1$, $L_c=100$ nm, $E_s=3.5$~eV, $g_N=0.4$~eV, and we have $k_\Vert=2\pi\sin\theta/\lambda$, where $\lambda$ is the wavelength of the cavity photon (here 400 nm).
  }
  \label{fig:hopfield}
\end{figure}

The new Hopfield coefficients satisfy $|\alpha_\pm^{(2)}|^2=|\gamma_\mp^{(2)}|^2$ and $|\alpha_0^{(2)}|^2=|\gamma_0^{(2)}|^2=|\beta_\pm^{(2)}|^2$, and they are plotted in Fig.~\ref{fig:hopfield}. From Fig.~\ref{fig:hopfield} we see that, with the chosen parameters, the singlet and cavity mode energies are in resonance at $\theta\approx32^\circ$. Here $|\alpha^{(1)}|^2=|\beta^{(1)}|^2=0.5$, i.e., both the LP and UP are half excitonic, half photonic, but perhaps less intuitively, also $2|\alpha_\pm^{(2)}|^2=|\beta_\pm^{(2)}|^2=0.5$ and $|\beta_0^{(2)}|^2=0$. It is also interesting to notice how the different coefficients behave in different subspaces. For example, while the fully excitonic/photonic coefficients of $|P_+\rangle$ (red/brown) and $|\psi_+^{(2)}\rangle$ (purple/green) follow the same trend with nearly constant difference, the half-excitonic-half-photonic coefficients $|\beta_l^{(2)}|^2$ behave in an independent way. These are not only interesting observations from a fundamental point of view but also important for optimizing specifically excitonic or photonic processes. Still, we shall simplify the analysis by restricting ourselves to $k_\Vert=\theta=0$ (i.e., normal incidence) in the rest of the article.

The remaining two-excitation eigenstates were recently derived in~\cite{two-exc-eigen}, and they are
\begin{align}
|\phi_{k,+}^{(2)}\rangle&=\beta^{(1)}\sum_{m<n}c_{mn}^{(k)}|S_mS_n\rangle\otimes|0\rangle+\alpha^{(1)}\hat{a}^\dagger|D_k\rangle,\\
|\phi_{k,-}^{(2)}\rangle&=\alpha^{(1)}\sum_{m<n}c_{mn}^{(k)}|S_mS_n\rangle\otimes|0\rangle-\beta^{(1)}\hat{a}^\dagger|D_k\rangle,\\
|\varphi_{kl}^{(2)}\rangle&=\sum_{m<n}c_{mn}^{(kl)}|S_mS_n\rangle\otimes|0\rangle,
\end{align}
with the respective eigenvalues of $E_s+E_+$, $E_s+E_-$, $2E_s$ and their multiplicities of $N-1$, $N-1$, $N(N-3)/2$. Here, $c_{mn}^{(k)}=\frac{2e^{i\pi(k-1)\frac{m+n}{N}}}{\sqrt{N(N-2)}}\cos\Big(\pi(k-1)\frac{m-n}{N}\Big)$ whereas the amplitudes $c_{mn}^{(kl)}$ have highly nontrivial forms that are not particularly enlightening to present here. All the eigenenergies both in the single- and two-excitation subspaces (1E and 2E) are sketched in Fig.~\ref{fig:schematic}a.

It might be tempting to write even higher-order eigenvalues as similar mixtures of $E_s$, $E_c$, and $E_\pm$. However, the higher the dimension $d$ of the system, the more erroneous the required approximation $N-d+1\approx N$ becomes, among other complications. Furthermore, the two subspaces considered are enough for the purposes of this article.

\subsection*{Enhancing RISC in and beyond the single-excitation subspace}

Since we are assuming weak singlet-triplet coupling, we can treat triplet-to-singlet transitions (i.e., RISC and TTA) perturbatively and, using Fermi's golden rule, write the transition rates as~\cite{Forrest2020}
\begin{equation}
k_{i\to f}=\frac{2\pi}{\hbar}|\langle f|H_{int}|i\rangle|^2\rho(E_{if}).
\label{eq:FGR}
\end{equation}
Here, $H_{int}$ and $\rho(E_{if})$ are the interaction Hamiltonian and joint density of states of the initial and final wavefunctions, respectively. In the case of bare-film RISC, $i=T_n$, $f=S_n$, and the interaction Hamiltonian reads $H_{int}=g_{st}\sum_n(|S_n\rangle\langle T_n|+|T_n\rangle\langle S_n|)$.

Electron transfer rates can be written alternatively by using the celebrated Marcus theory. Assuming weak phonon coupling, the rates are given by~\cite{Jortner1976}
\begin{align}
    k_{i\to f}&=k(\lambda,\Delta E)\\
    &\approx\frac{g_{if}^2}{\hbar}\sqrt{\frac{\pi}{\lambda k_BT}}\exp\Big[-\frac{(\lambda+\Delta E)^2}{4\lambda k_BT}\Big],
    \label{eq:Marcus}
\end{align}
where $g_{if}$ is the coupling strength, $\lambda$ the reorganization energy, $\Delta E$ the change in free energy, and $T$ the temperature. The bare-film RISC rate therefore becomes $k_{RISC}^s=k(\lambda_s,E_s-E_t)$ with $g_{if}=g_{st}$.

Let us proceed with the polaritonic case. According to Hund's law, the first-order triplet state always has lower energy than the first-order singlet state~\cite{Yu2021}. Therefore, it is the LP that can be expected to enhance RISC. The rate of triplet-to-LP RISC can be obtained by applying Eq.~\eqref{eq:FGR}--\eqref{eq:Marcus} to LP. Assuming that the cavity has no effect on $g_{st}$ nor the functional form of $\rho(E_{if})$, we get
\begin{align}
k_{RISC}^-&=k_{T_n\to P_-}\\
&=\frac{2\pi}{\hbar}|\langle P_-|H_{int}|T_n\rangle|^2\rho(E_{t-})\\
&=\frac{2\pi}{\hbar}\Big|g_{st}\frac{\beta^{(1)*}}{\sqrt{N}}\Big|^2\rho(E_{t-})\\
&=\frac{|\beta^{(1)}|^2}{N}k(\lambda_-,E_--E_t).
\label{eq:LP-RISC-rate}
\end{align}
The collective nature of polaritons---manifested here as the denominator $N$---dramatically hinders the triplet-to-LP RISC. However, the initial energy gap $E_s-E_t$ has the opposite effect; With great $E_s-E_t$, RISC can be enhanced with strong enough coupling and low enough LP energy.

Fig.~\ref{fig:schematic}b shows the singlet and triplet energies at which RISC can be enhanced with strong coupling (and our example parameters). The reorganization energies were chosen according to~\cite{Eizner2019}, $\lambda_s=0.10$~eV and $\lambda_-=0.79$~eV. In the colored region, the heat map gives the value of $g_N$ at which the LP-RISC rate equals the bare-film rate, $k_\textit{RISC}^-=k_\textit{RISC}^s$. Note that according to this definition of enhanced RISC, we are actually neglecting the UP and exciton reservoir. This is justified because the UP is far from resonance with the triplets, and the exciton reservoir is dark and therefore against our ultimate goal of getting more light from OLEDs. Nevertheless, it can be estimated that when $k_\textit{RISC}^-=k_\textit{RISC}^s$, the \textit{total} RISC rate---in terms of triplet depopulation---is actually \textit{twice} the original (see Supplementary Note 1).

The colored region in Fig.~\ref{fig:schematic}b displays both very small and very large coupling strengths. In the blue region, the initial RISC rates are negligible, which means that both the required coupling strengths and \textit{absolute} enhancements are also very small. Furthermore, polaritons are not stable in the blue region, because $g_N$ must be greater than losses for polaritons to form~\cite{Bhuyan2023}. But since Fig.~\ref{fig:schematic}b only shows the minimum requirements, both stable polaritons and meaningful absolute enhancements can be achieved with larger coupling strengths. This is illustrated in Supplementary Figure 1a, where we have plotted the ratios $k_\textit{RISC}^-/k_\textit{RISC}^s$ as functions of $g_N$. But while stronger couplings give higher rates, more careful analysis is required at very high coupling strengths, since the rotating-wave approximation might not be valid anymore.

In the white region of Fig.~\ref{fig:schematic}b, we have $k_\textit{RISC}^-<k_\textit{RISC}^s$, which means that RISC cannot be enhanced. In fact, the white region corresponds to singlet-triplet splittings typical for efficient RISC materials, revealing that there is no prospect to enhance RISC in already efficient systems. However, here we stopped the numerical run at $g_N=1$~eV; Higher coupling strengths have not been reported in related literature, and achieving higher coupling strengths would be pragmatically very challenging. It would require absorbers that combine a very high dipole moment, very small size, and ability to pack densely and uniformly along the electric field polarization. 

The role of reorganization energies in RISC is substantial. While they were fixed here, we have considered different values in the Supplementary Information. In Supplementary Figure 2, we have varied them in the intervals $\lambda_s/\text{eV}\in[0.1,0.5]$ and $\lambda_-/\text{eV}\in[0.6,1.0]$. $\lambda_->\lambda_s$ because polaritons are partially in the electronic ground state, which is further away from the triplet states than the excited singlet states. Supplementary Figure 2 shows that the greater the difference between $\lambda_-$ and $\lambda_s$, the higher the coupling strength required to enhance RISC.

A few more remarks are in order. While the rate $k_{RISC}^-$ is applicable in the presence of any number of triplets, here we implicitly assumed that there are no excitations in the polariton manifold before RISC---or that the single-excitation polaritons depopulate fast enough before the consecutive RISC processes; Typical fluorescence rates of organic molecules are of the order of $10^8$~s$^{-1}$, while high performance materials have displayed RISC rates of $10^4$--$10^6$~s$^{-1}$~\cite{Forrest2020}. If this is not the case, we go from the single-excitation subspace to the two-excitation subspace, where the closest state in energy is $|\psi_-^{(2)}\rangle$. However, since $|\psi_-^{(2)}\rangle$ is nondegenerate in energy and there are $N(N+1)/2$ other states, the reciprocal scaling of RISC can be expected to become even worse. Hence, the ratio $k_\textit{RISC}^-/k_\textit{RISC}^s$ should actually be interpreted as the \textit{upper bound} of relative RISC enhancement, in terms of excitation number.

When bringing LP closer to the first-order triplet state, we inevitably enhance inter-system crossing (ISC) as well. This is against our goal: to convert the slow triplets to fast singlets (or polaritons). Yet, while RISC is weighted by the long-lived triplet population, ISC is weighted by the much smaller LP population. This is an important point we want to highlight, since high RISC/ISC ratio \textit{can} be achieved if the triplets have long enough lifetimes and the LP empties quickly enough. Finally, even though we have focused on LP, we might be able to harvest higher-order triplets with UP and ``hot RISC" as well~\cite{Lin2021}.

\subsection*{Enhancing TTA}

Because there are three triplet states, under the simplest scheme two of them can combine in nine different ways when colliding. 1/9 of the encounter complexes are of the singlet character---denoted here by $\mathbf{S}$---that can relax to the singlet branch, while 3/9 and 5/9 are of the irrelevant triplet and quintet character $\mathbf{T}$ and $\mathbf{Q}$, respectively~\cite{Ye2021}. Because the triplets are uncoupled from the cavity, their collision rate can be expected to remain constant. Therefore, we focus on the relaxation of encounter complexes and whether it can be enhanced with polaritons.

The exact dynamics of encounter complexes is a subject of active research, including whether and how the $\mathbf{Q}$ states impact the spin statistics of TTA~\cite{Andrewa1,Andrewa2,Andrewa3} and the optical accessibility of the encounter complex itself~\cite{Andrewb,Andrewc}. In systems where twice the triplet energy is close to the bright singlet, configuration interactions can substantially complicate the wavefunctions and shift the energies of the triplet pairs relative to two triplets~\cite{Andrewd}. Moreover, in solid systems it can be essential to factor in the role of longer-range encounter complexes where weaker exchange coupling between triplets results in spin mixing between $\mathbf{S}$ and $\mathbf{Q}$ states that enriches the simple picture laid out above~\cite{Andrewe}. The detailed evolution of these pair states is strongly material-dependent and poorly understood~\cite{Andrewf}, but aside from specially designed systems~\cite{Berghuis2019,Ye2021,Andrewb,Andrewc,Andrewe}, these encounter complexes should be fleetingly short-lived. In particular, in typical OLED materials the very large energy gap between singlet and triplet pair should minimize the impact of these complications. Thus, it is sufficient for our purposes here to use, similarly to RISC, non-zero coupling strength of singlets and encounter complexes $g_{sc}$ that does not change within the cavity.

We take the interaction Hamiltonian of the $\mathbf{S}$ states and singlets to be of the form $H_{int}=g_{sc}\sum_{m<n}\big[(|S_m\rangle+|S_n\rangle)\langle\mathbf{S}_{m,n}|+|\mathbf{S}_{m,n}\rangle(\langle S_m|+\langle S_n|)\big]$ (cf. Ref.~\cite{Nakano2016}). Again, we apply both the Marcus theory and Fermi's golden rule. Assuming reorganization and free energies twice as high as those with RISC, we obtain
\begin{equation}
    k_{TTA}^s=k(2\lambda_s,E_s-2E_t)
    \label{eq:bare-TTA}
\end{equation}
in the bare-film case and
\begin{align}
    k_{TTA}^+&=\frac{4|\alpha^{(1)}|^2}{N}k(2\lambda_+,E_+-2E_t),
    \label{eq:UP-TTA}\\
    k_{TTA}^-&=\frac{4|\beta^{(1)}|^2}{N}k(2\lambda_-,E_--2E_t)
    \label{eq:LP-TTA}
\end{align}
in the polariton cases. The factor of 4 in the numerators of $k_{TTA}^\pm$ is explained by both possible singlet states contributing to the same polariton, effectively doubling its probability amplitude. Still, because polaritons are collective states and TTA an intermolecular process, one might have anticipated even more prominent enhancement. In fact, the situation might be different with non-negligible triplet TDM. Consider sparsely excited triplets, i.e., triplets outside their capture radii and therefore not promoting TTA. As the triplet polaritons would consist of all the possible permutations of triplet-occupied molecular sites, some of them might be within the capture radii and contribute to TTA. On the other hand, we would also have the competing process of strong coupling ``separating" triplets. This idea is discussed later in more detail with singlet-singlet annihilation (SSA).

Eqs.~\eqref{eq:bare-TTA}--\eqref{eq:LP-TTA} might imply the stability of encounter complexes: the bigger the free-energy difference, the slower they relax to singlets. There are, however, competing channels that effectively reduce TTA to the singlet branch. In fact, we suspect that the doubled reorganization energies pose a severe penalty against TTA in the presence of these other channels. In particular, triplet encounter complexes can relax back to free triplets and decay non-radiatively~\cite{Yong2017}. Nevertheless, since both can be assumed ubiquitous and unaffected by the cavity, they do not affect the bare-film/cavity comparison and can therefore be omitted.

Fig.~\ref{fig:schematic}c shows the singlet and triplet energies at which $\mathbf{S}$-to-UP TTA can be enhanced with strong coupling. Again, we have $k_\textit{TTA}^+=k_\textit{TTA}^s$ in the colored region and $k_\textit{TTA}^+<k_\textit{TTA}^s$ in the white region, i.e., the region of already efficient TTA (see, e.g., Ref.~\cite{Suzuki2014}). LP-TTA has been considered in the Supplementary Information, where we have plotted the ratios $k_{TTA}^\pm/k_{TTA}^s$ and performed a more detailed scan of reorganization energies (Supplementary Figures 3 and 4). Comparing Figs.~\ref{fig:schematic}b and c, one can see that the boundary between enhancement and no enhancement is twice as steep with TTA as with RISC, which can be explained by the doubled triplet energies. It is also very interesting and important to notice that either RISC or TTA---or \textit{both}---can always be enhanced, at least relatively.

\subsection*{Effects of $N$}

The number of molecules plays many roles in polaritonic OLEDs. In particular, there are two competing effects of $N$ with single-excitation polaritons. On one hand, we can move the polariton states and close the free-energy gaps $\Delta E$ in the Marcus rates by increasing $g_N$ which, in a fixed mode volume $V$ and molecular TDM, means increasing $N$. On the other hand, the factor of $1/N$ strongly dilutes the polaritonic rates as shown in the previous two sections. The number (or density) of molecules also affects TTA, SSA, and other annihilation processes, but with triplet-triplet collision rate there should be no connection to strong coupling due to negligible TDM. $N$-dependent SSA, on the other hand, will be discussed shortly.

\begin{figure}[t!]
  \includegraphics[width=\linewidth]{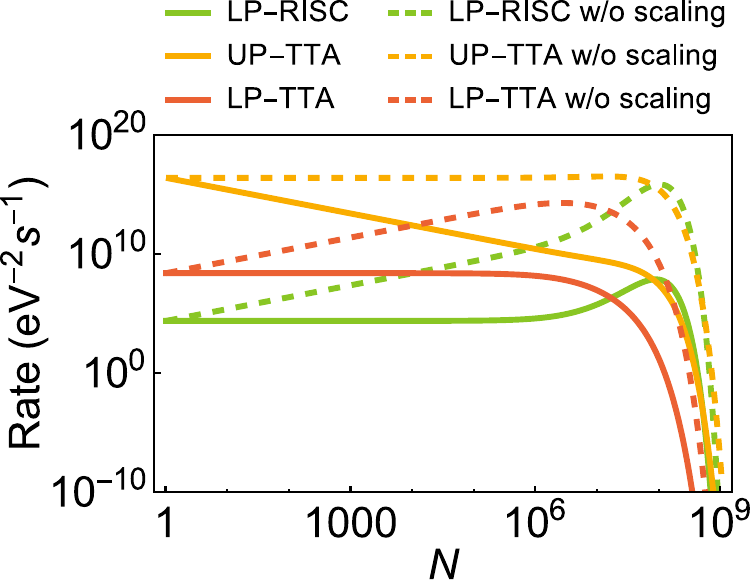}
  \caption{
  \textbf{ \textit{N}-dependent rates of 3DPA3CN.} The rates are shown in units of $g_{st}^2$ (RISC) or $g_{sc}^2$ (TTA). The parameters are $E_s=2.92$~eV, $E_t=2.41$~eV, $E_c=2.56$~eV, $\lambda_s=0.10$~eV, $\lambda_\pm=0.79$~eV, and $T=293$~K.
  }
  \label{fig:N-dep-rates}
\end{figure}

We have plotted the polaritonic rates of the molecule 1,3,5-tris(4-(diphenylamino)phenyl)-2,4,6-tricyanobenzen (3DPA3CN) in Fig.~\ref{fig:N-dep-rates} both with and without the inverse scaling. With the parameters reported in Ref.~\cite{Eizner2019}, the coupling strength can be expressed as $g_N=\sqrt{N}\times112.5\times10^{-6}$ eV. Comparing the solid (correct rates) and dashed (no inverse scaling) curves of the same color, we can clearly see how $1/N$ hinders, and in the case of UP-TTA, even weakens the rates. The critical value of $N$ is the most prominent with RISC; The rates increase in a Gaussian fashion just before $N\approx10^8$, after which $E_-$ surpasses $E_t-\lambda_-$ and the rates quickly die. Finally, although the rates of UP-TTA and LP-TTA are orders of magnitude larger than LP-RISC, one should keep in mind that the triplet excitons should first collide, and the values of $N$ in Fig.~\ref{fig:N-dep-rates} are relatively small in planar microcavities. In fact, no enhancement of TTA was reported in~\cite{Eizner2019}, where the number of coupled molecules was estimated to be only $4\times10^6$.

\subsection*{Reducing SSA}

When two singlet excitons interact, they are promoted to a higher-order excited singlet which then either relaxes back to the first-order singlets or ground state---releasing heat at the same time---or breaks into free charge carriers~\cite{King2007}. This may lead to efficiency roll-off and device degradation~\cite{Gather2011}. However, with strong coupling it should be possible to ``separate" such close singlet excitons, as the formed polaritons also consist of \textit{distant} singlets.

\begin{figure*}[t!]
  \includegraphics[width=\textwidth]{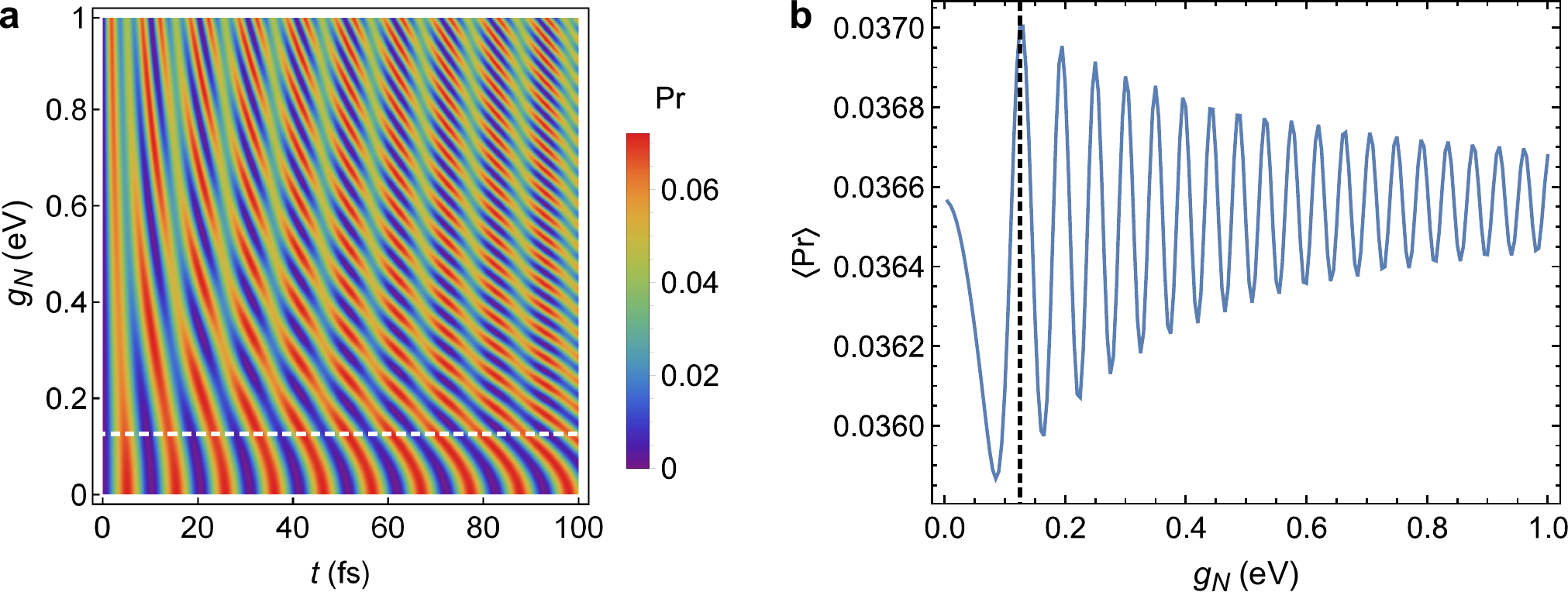}
  \caption{
  \textbf{ Reducing SSA. a, }Probability of singlet pairs outside the capture radius (i.e., probability of \textit{not} getting SSA) as a function of coupling strength and time. \textbf{b, }Arithmetic mean of the ``no SSA" probability as a function of $g_N$. The used example parameters are $N=100$, $E_s=3.5$ eV, $n_\textit{eff}=2$, $m=1$, $L_c=100$ nm, and $k_\Vert=0$.
  }
  \label{fig:ssa}
\end{figure*}

Let us clarify this idea by deriving a na\"{i}ve estimate for the ``no SSA" probability. Consider the initial state $|S_mS_n\rangle\otimes|0\rangle$, where the two singlets are close to each other and about to annihilate. Focusing solely on the unitary dynamics induced by $U(t)=\exp(-iHt/\hbar)$, the probability to find the singlets at the same sites at time $t$ is $P_{m,n}(t)=|\langle S_mS_n|U(t)|S_mS_n\rangle|^2$ which, after some algebra, can be written as
\begin{equation}
\begin{split}
    P_{m,n}(t)=&\Bigg|\frac{2}{N(N-1)}\Big(|\alpha_+^{(2)}|^2e^{-i\frac{2E_+t}{\hbar}}+|\alpha_0^{(2)}|^2e^{-i\frac{(E_s+E_c)t}{\hbar}}\\
    &+|\alpha_-^{(2)}|^2e^{-i\frac{2E_-t}{\hbar}}\Big)+\frac{2}{N}\Big(|\beta^{(1)}|^2e^{-i\frac{(E_s+E_+)t}{\hbar}}\\
    &+|\alpha^{(1)}|^2e^{-i\frac{(E_s+E_-)t}{\hbar}}\Big)+\frac{N-3}{N-1}e^{-i\frac{2E_st}{\hbar}}\Bigg|^2.
\end{split}
\end{equation}
Note that this equation only holds for very short times, until incoherent effects take place. If we further assume a cubic-centered lattice of molecular sites and the annihilation of nearest neighbours only, the probability to have further-apart singlets becomes
\begin{equation}
    \text{Pr}(t)=\frac{N^2-9N}{N^2-N-2}\big[1-P_{m,n}(t)\big].
    \label{eq:no-SSA}
\end{equation}

Fig.~\ref{fig:ssa}a shows the ``no SSA" probability~\eqref{eq:no-SSA} as a function of both $g_N$ and $t$. In the Tavis-Cummings model, emitters interact collectively with the cavity mode. The initial state $|S_mS_n\rangle$ is not a symmetric state under permutation of emitters. Symmetric states, such as Dicke states, are typically more efficient in exchanging excitations with the cavity mode. Since $|S_mS_n\rangle$ is not symmetric, the coupling to the cavity mode is less efficient, which explains the low probabilities. Note also that here we only considered unitary dynamics. Incoherent cavity losses, described by the annihilation operator $\hat{a}$, might result in more symmetric transitions between subspaces and therefore bigger ``no SSA" probabilities.

Interestingly, when taking the average of ``no SSA" probability over multiple instances, an optimum value of $g_N$ can be observed from Fig.~\ref{fig:ssa}b ($\sim125$ meV). Here, the timespan was from 0 to 100 fs in timesteps of 0.1 fs. Fig.~\ref{fig:ssa} also demonstrates that, even though electrical excitation may be spatially confined in a thin recombination zone, with strong coupling singlets can be formed outside of it. OLEDs with polariton-improved operational lifetimes were recently reported in~\cite{Zhao2024}, to which our analysis ultimately provides an alternative (or complementary) explanation.

\setlength{\arrayrulewidth}{0pt}

\setlength{\tabcolsep}{0pt}

\renewcommand{\arraystretch}{1.2}

\captionsetup{labelformat=nctable,labelsep=vert,justification=raggedright}

\begin{table*}
  \caption{\protect{\textbf{ RISC and TTA rates.}} The rates are given in units of $g_{st}^2$ or $g_{sc}^2$. The bare-film intervals were obtained with $\lambda_s/\text{eV}\in[0.1,0.5]$ and the polaritonic intervals with $\lambda_\pm/\text{eV}\in[0.6,1.0]$ except for 3DPA3CN, for which we used the reported values of $\lambda_s=0.10$ eV and $\lambda_\pm\approx0.79$ eV. Note also the scaling of the polaritonic rates; The actual rates are obtained by dividing with $N$, the number of molecules.}
  \begin{tabular}{ c c c c c c }
  \rowcolor{orange!30}
    \hspace{7pt}\textbf{Molecule}\hspace{7pt} &
    \hspace{7pt}$\textit{\textbf{k}}_\textit{\textbf{RISC}}^\textit{\textbf{s}}\mathbf{/}\textit{\textbf{g}}_\textit{\textbf{st}}^2$\hspace{7pt} &
    \hspace{7pt}$\textit{\textbf{Nk}}_\textit{\textbf{RISC}}^\mathbf{-}\mathbf{/}\textit{\textbf{g}}_\textit{\textbf{st}}^2$\hspace{7pt} &
    \hspace{7pt}$\textit{\textbf{k}}_\textit{\textbf{TTA}}^\textit{\textbf{s}}\mathbf{/}\textit{\textbf{g}}_\textit{\textbf{sc}}^2$\hspace{7pt} &
    \hspace{7pt}$\textit{\textbf{Nk}}_\textit{\textbf{TTA}}^\mathbf{+}\mathbf{/}\textit{\textbf{g}}_\textit{\textbf{sc}}^2$\hspace{7pt} &
    \hspace{7pt}$\textit{\textbf{Nk}}_\textit{\textbf{TTA}}^\mathbf{-}\mathbf{/}\textit{\textbf{g}}_\textit{\textbf{sc}}^2$\hspace{7pt} \\
    Erythrosine B~\cite{Stranius2018} & $10^{9}$--$10^{11}$ & $10^{10}$--$10^{11}$ & $10^{-24}$--$10^{15}$ & $10^{15}$--$10^{16}$ & $10^{16}$--$10^{16}$ \\
    \rowcolor{orange!30}
    DABNA-2~\cite{Yu2021} & $10^{11}$--$10^{13}$ & $10^{12}$--$10^{14}$ & $10^{-164}$--$10^{1}$ & $10^{10}$--$10^{16}$ & $10^{0}$--$10^{13}$ \\
    Tetracene~\cite{Berghuis2019} & $10^{-44}$--$10^{-6}$ & $10^{-2}$--$10^{-1}$ & $10^{13}$--$10^{16}$ & $10^{8}$--$10^{12}$ & $10^{10}$--$10^{13}$ \\
    \rowcolor{orange!30}
    DPP(PhCl)$_2$~\cite{Ye2021} & $10^{-51}$--$10^{-7}$ & $10^{-2}$--$10^{-1}$ & $10^{11}$--$10^{15}$ & $10^{4}$--$10^{7}$ & $10^{8}$--$10^{12}$ \\
    3DPA3CN~\cite{Eizner2019} & $10^{15}$ & $10^{12}$ & $10^{-79}$ & $10^{16}$ & $10^{14}$ \\
    \rowcolor{orange!30}
    TDAF~\cite{Nanoph2023} & $10^{-21}$--$10^{1}$ & $10^{8}$--$10^{10}$ & $10^{-22}$--$10^{15}$ & $10^{14}$--$10^{16}$ & $10^{13}$--$10^{16}$ \\
  \end{tabular}
  \label{tbl:Table1}
\end{table*}

\subsection*{Comparing the model to experimental data}

We calculated the RISC and TTA rates for six molecules previously studied under strong coupling, using the same parameters as in the original works (given in Supplementary Table 1). To achieve in-depth understanding on the role of reorganization energies, we scanned the intervals $\lambda_s/\text{eV}\in[0.1,0.5]$ and $\lambda_\pm/\text{eV}\in[0.6,1.0]$ (except for 3DPA3CN). Table~\ref{tbl:Table1} shows the molecules and their minimum and maximum RISC and TTA rates at these intervals. The bare-film intervals can be notably large, whereas the role of reorganization energies becomes smaller in the cavity cases. That is, the closer the reorganization energies are to zero (as with the bare-film case), the more their precise values matter. Because we did not consider \textit{all} the possible OLED processes, some theoretical rates might substantially diverge from the reported experimental values.

The first experimental study of the impact of polaritonic states on RISC was realized in Ref.~\cite{Stranius2018} with Erythrosine B, a material exhibiting both delayed fluorescence and phosphorescence. The authors reported direct transition between molecular-centered and polaritonic states when bringing the LP closer to the first-order triplet. Enhanced RISC was later reported for also 9-([1,1-biphenyl]-3-yl)-N,N,5,11-tetraphenyl-5,9-dihydro-5,9-diaza-13b-boranaphtho[3,2,1-de]anthracen-3-amine (DABNA-2)~\cite{Yu2021}. While our model predicts enhanced RISC for both molecules with very small $N$, for larger values of $N$ it actually favors TTA. However, if there are enough molecules for TTA to occur, then the harmful, non-emitting SSA may also occur, which we have shown strong coupling to reduce. In other words, enhanced emission can be due to reduced SSA as well. Furthermore, there are other mechanisms behind enhanced emission not included in our model, e.g., Purcell enhancement~\cite{Vahala2003,Peng2015} and polariton-induced spectral filtering with refocused emission intensity~\cite{Daskalakis2019,Khazanov2023}.

With tetracene and DPP(PhCl)$_2$ we have the exact opposite situation; While enhanced TTA was reported~\cite{Berghuis2019,Ye2021}, our model favors RISC. The discrepancy with tetracene may be explained by the fact that in~\cite{Berghuis2019} the physical system was slightly different, utilizing nanoparticle arrays exhibiting surface lattice resonances which have smaller mode volumes compared with planar microcavities. In Ref.~\cite{Ye2021}, an endothermic TTA process was turned into exothermic by lowering the LP below the $\mathbf{S}$ encounter complex. The theoretical model, however, was based on simple Arrhenius rates omitting reorganization energies. While Purcell enhancement was ruled out in all the previous demonstrations, it would have been worthwhile to consider experiments that disregard enhanced emission due to polariton filtering or reduced SSA. We note that a further complication is that all the reported processes correspond to long-lived and mixed-spin encounter complexes, and it may be necessary to incorporate a fuller description of triplet-pair evolution to capture their behavior.

In the above four cases, where either RISC or TTA was reportedly enhanced, the phonon couplings that we assumed weak might have actually played a bigger role. However, even when taking the phonon couplings better into consideration and comparing the upper bounds of the rates---obtained by omitting the Gaussian damping in Eq.~\eqref{eq:Marcus}---our model alone did not explain the experimental findings.

Finally, no enhancement was reported for either 3DPA3CN~\cite{Eizner2019} or 2,7-Bis[9,9-di(4-methylphenyl)-fluoren-2-yl]-9,9-di(4-methylphenyl)fluorene (TDAF)~\cite{Nanoph2023}  even though our model predicts major improvement of TTA and, depending on $\lambda_s$, RISC for TDAF---in fact, TDAF is the only molecule with both RISC and TTA potentially enhanced. But as already mentioned, the number of molecules did not support TTA with 3DPA3CN. As for TDAF, the number of molecules may very well explain the nonexistent RISC and TTA, yet in~\cite{Nanoph2023} both might have been (also) smeared by delayed fluorescence from trapped charge carriers. This highlights the importance of ruling out all sources of delayed fluorescence, whether enhanced or not. Summarizing, although we have taken significant steps towards a model connecting all the experimental findings, both more comprehensive theories and experiments are still needed to verify if RISC and TTA can be enhanced in realistic devices.

\section*{Discussion}

In this article, we derived polaritonic RISC and TTA rates in the presence of weak phonon coupling. Comparing with the bare-film case, we were able to identify the parameter spaces of enhanced RISC and TTA, both of which heavily depend on the singlet and triplet free energies, reorganization energies---a factor often omitted from prior works---and number of coupled molecules. Hence, our results help in designing next-generation OLEDs without the need for trial and error.

To achieve polaritonic enhancements, the modified energy spacings must counteract both the unfavorable reorganization energies and the effects of large $N$. Since both can be significant, the change in free energy---and therefore the initial singlet-triplet gap being compared---must also be large. This implies that the original bare-film processes need to be very inefficient to begin with. Processes that are already efficient cannot be further enhanced by polaritons.

Because the inverse-scaling problem is expected to hinder strong polariton-induced dynamics in the emerging field of polariton organic optoelectronics, enhanced RISC and TTA can be somewhat challenging to distinguish from Purcell enhancement and polariton filtering. To carefully rule out such alternative sources of enhanced emission in future experiments, both more sophisticated models and experiments are still needed. Such models would, e.g., take into account all the couplings and processes we omitted. Furthermore, one should really consider a continuum of cavity modes as in~\cite{Lydick2024}.

Even though it might appear an elusive goal to fully harness triplets with polaritonic RISC and TTA, strong coupling might influence OLEDs in other, perhaps more surprising ways. Strong coupling can, in a sense, redistribute excitons. This can benefit fluorescent materials by either enhancing TTA or reducing SSA. While further research is needed to fully understand these fascinating new directions, strong coupling clearly holds tremendous potential for next-generation OLEDs. In general, our work helps to better understand the rich dynamics occurring in polariton OLEDs and paves the way for more advanced hybrid light-matter technologies.

\section*{Acknowledgments}

This project has received funding from the European Research Council (ERC) under the European Union’s Horizon 2020 research and innovation programme (grant agreement No. [948260]). The authors would like to thank Ahmed Abdelmagid for useful comments.

\section*{Author contributions}

O.S. and K.S.D. conceived the work. O.S. performed the theoretical analysis and wrote the article together with K.S.D. and A.J.M. K.L. oversaw the theoretical analysis and K.S.D. supervised the work. All authors discussed the results and contents of the article. 

\section*{Competing interests}

The authors declare no competing interests.

\bibliography{refs}

\end{document}